\documentclass[a4paper,10pt]{article}

\usepackage[utf8]{inputenc}
\usepackage{amsmath,amssymb}
\newtheorem{thm}{Theorem}[section]

\newtheorem{cor}[thm]{Corollary}
\usepackage[all]{xy}

\title{On vanishing theorems for Higgs bundles}
\author{S. A. H. Cardona\footnote{Electronic address: andres.holguin@cimat.mx}\\CIMAT A.C. - Via Jalisco S/N - 36240, Gto. - M\'exico}

\begin{document}

\maketitle

\begin{abstract}
We introduce the notion of Hermitian Higgs bundle as a natural generalization of the notion of Hermitian 
vector bundle and we study some vanishing theorems concerning Hermitian Higgs bundles when the base manifold 
is a compact complex manifold. We show that a first vanishing result, proved for these objects when the base 
manifold was K\"ahler, also holds when the manifold is compact complex. From this fact and some basic 
properties of Hermitian Higgs bundles, we conclude several results. In particular we show that, in analogy to 
the classical case, there are vanishing theorems for invariant sections of tensor products of Higgs bundles. 
Then, we prove that a Higgs bundle admits no nonzero invariant sections if there is a condition of 
negativity on the greatest eigenvalue of the Hitchin-Simpson mean curvature. Finally, we prove that invariant 
sections of certain tensor products of a weak Hermitian-Yang-Mills Higgs bundle are all 
parallel in the classical sense.\\

\end{abstract}

\section{Introduction}

As it is well known, in complex geometry one has some results on vanishing of holomorphic sections of a 
holomorphic vector bundle under certain negativity conditions on the Chern mean curvature of the bundle. These results, first proved by Bochner and Yano 
\cite{Yano-Bochner}, have been used by Gauduchon \cite{Gauduchon} and Kobayashi \cite{Kobayashi} to study some 
properties of Hermitian vector bundles over compact complex manifolds. In particular, Kobayashi used some of 
these properties to prove one direction of the Hitchin-Kobayashi correspondence for holomorphic 
vector bundles over compact K\"ahler manifolds; namely, Kobayashi proved the polystability of such a bundle if 
the bundle was Hermitian-Einstein. The other direction of this correspondence has been proved by Donaldson 
\cite{Donaldson-2}, \cite{Donaldson-3} when the base manifold was a compact complex projective manifold, and 
by Uhlenbeck and Yau \cite{Uhlenbeck-Yau} when the manifold was compact K\"ahler. The Hitchin-Kobayashi 
correspondence plays an important role in Complex Geometry and is the subject of much active research, it has 
been studied in detail by L\"ubke and Teleman \cite{Lubke} in the case of holomorphic vector bundles when the base manifold is compact complex, and 
has been extended to coherent sheaves over compact K\"ahler manifolds by Bando and Siu \cite{Bando-Siu}.\\

On the other hand, the notion of a Higgs bundle was introduced by Hitchin \cite{Hitchin} and Simpson \cite{Simpson}, \cite{Simpson 2}. 
They used this concept to construct an extension of the Hitchin-Kobayashi correspondence. In particular, in 
\cite{Hitchin} Hitchin used some vanishing theorems to prove that an irreducible Higgs bundle over a compact 
Riemann surface of zero degree is stable if and only if it satisfies the Hermitian-Yang-Mills condition. 
In \cite{Simpson}, Simpson extended the result of Hitchin for Higgs bundles over K\"ahler manifolds of 
arbitrary dimension and degree, which is indeed the Hitchin-Kobayashi correspondence for Higgs bundles. 
Following the ideas of Bando and Siu \cite{Bando-Siu}, Biswas and Schumacher \cite{Biswas-Schumacher} proved 
this correspondence also for Higgs sheaves. Now, Bruzzo and Gra\~na Otero \cite{Bruzzo-Granha} proved one of 
these vanishing results for Higgs bundles over compact K\"ahler manifolds, and they used it to proved that a 
Higgs bundle is semistable if it admits an approximate Hermitian-Yang-Mills metric. Finally, Seaman in 
\cite{Seaman} studied some particular examples of Higgs bundles, which arise in a natural way from bundles of 
holomorphic forms, and he proved some vanishing theorems for holomorphic forms on these bundles. \\

This article is organized as follows, in the first section we introduce the notion of a Hermitian Higgs 
bundle over a compact Hermitian manifold and we make some comments about the basic properties of these objects. 
In particular, we review the notion of invariant section of a Hermitian Higgs bundle. For Hermitian Higgs 
bundles we can apply the same operations that are commonly applied to Hermitian vector bundles, 
and hence make sense to study the same notions that are introduced in Complex Geometry. In this section we 
review the notion of Hitchin-Simpson curvature and we show that we get a formula for the 
corresponding mean curvature, which is indeed similar to the classical one. We also introduce the concept of 
weak Hermitian-Yang-Mills metric, as a natural generalization of the concept of weak Hermitian-Einstein 
metric introduced by Kobayashi \cite{Kobayashi}. In the final part of this section we review the classical 
Weitzenb\"ock formula, a key result that can be used also for Hermitian Higgs bundles.  \\

In the second section we prove some Bochner's vanishing theorems for Hermitian Higgs bundles over compact 
complex manifolds. In the first part we show that if the Hitchin-Simpson mean curvature of a Hermitian Higgs 
bundle is seminegative definite everywhere, every invariant section is parallel in the classical sense, i.e., 
it is parallel with respect to the Chern connection. If moreover, the Hitchin-Simpson mean curvature 
is negative definite at some point, then there are no nonzero 
invariant sections for such a bundle. This result has been proved in \cite{Bruzzo-Granha}, when 
the base manifold is K\"ahler. Here we modify their proof to cover also the general case. Then we show that, 
in analogy to the classical case, there is also a vanishing theorem for tensor products of Hermitian Higgs 
bundles and from this result we get some corollaries. Next, we prove the main theorem of this article. Namely, 
we prove that if the eigenvalues of the Hitchin-Simpson mean curvature of a Hermitian Higgs bundle satisfy 
certain negativity condition, then such a bundle admits no nonzero invariant sections. This result is again 
an extension of a classical result for Hermitian vector bundles. Finally, we prove that if a Hermitian Higgs 
bundle satisfies the Hermitian-Yang-Mills condition, then on certain tensor products of this bundle, every 
invariant section is parallel in the classical sense. \\

\section{Hermitian Higgs bundles}

We start with some basic definitions. Let $X$ be a compact complex manifold and denote by $\Omega_{X}^{1}$ the 
cotangent bundle to it. Following \cite{Hitchin} and \cite{Simpson}, a Higgs bundle ${\mathfrak E}$ over $X$ 
is a holomorphic vector bundle $E$ over $X$ together with a map $\phi : E\rightarrow E\otimes\Omega_{X}^{1}$ 
such that $\phi\wedge\phi : E\rightarrow E\otimes\Omega_{X}^{2}$ vanishes. The map $\phi$ is called the Higgs 
field of ${\mathfrak E}$. On Higgs bundles we can apply the same operations that we apply to holomorphic 
bundles. In particular, the dual of a Higgs bundle is again a Higgs bundle, and tensor products of Higgs 
bundles are Higgs bundles. If ${\mathfrak E}$ is a Higgs bundle we denote its dual by ${\mathfrak E}^{*}$, 
and if ${\mathfrak E}_{1}$ and ${\mathfrak E}_{2}$ are Higgs bundles over $X$, we denote by 
${\mathfrak E}_{1}\otimes{\mathfrak E}_{2}$ its tensor product. For further details about 
these basic properties see for instance \cite{Cardona} and \cite{Cardona 2}. Now, in order to establish 
the vanishing theorems we need to use the notion of invariant section of a Higgs bundle. Following 
\cite{Bruzzo-Granha}, we say that a section $s$ of a Higgs bundle ${\mathfrak E}$ is $\phi$-invariant if 
$\phi(s)=s\otimes\lambda$ for some holomorphic 1-form $\lambda$ on $X$.\\

A compact Hermitian manifold is a pair $(X,g)$, where $X$ is a compact complex manifold and $g$ is 
a Hermitian metric on $X$; a Hermitian Higgs bundle over $(X,g)$ is a pair\footnote{Notice that since a Higgs 
bundle ${\mathfrak E}$ is indeed a pair $(E,\phi)$, a Hermitian Higgs bundle can be seen also as a triple 
$(E,\phi,h)$.} $({\mathfrak E},h)$ where ${\mathfrak E}$ is a Higgs bundle over $X$ and $h$ is a Hermitian 
metric on $E$. In other words, a Hermitian Higgs bundle is just a Hermitian vector bundle (in the sense of 
Kobayashi \cite{Kobayashi}) such that the corresponding holomorphic bundle is a Higgs bundle. By definition, 
an invariant section of a Hermitian Higgs bundle $({\mathfrak E},h)$ is just an invariant section of 
${\mathfrak E}$. For a pair $({\mathfrak E},h)$ its dual Hermitian Higgs bundle is the pair 
$({\mathfrak E}^{*},h^{*})$ with $h^{*}$ the usual metric on the holomorphic bundle $E^{*}$ induced by $h$. 
Given $({\mathfrak E}_{1},h_{1})$ and $({\mathfrak E}_{2},h_{2})$ Hermitian Higgs bundles over $(X,g)$, then 
$({\mathfrak E}_{1}\otimes{\mathfrak E}_{2},h_{1}\otimes h_{2})$ is a Hermitian Higgs bundle over $(X,g)$, 
where $h_{1}\otimes h_{2}$ is the usual Hermitian metric on $E_{1}\otimes E_{2}$.\\

Let $({\mathfrak E},h)$ be a Hermitian Higgs bundle over $(X,g)$, as it is well known, there exists a unique 
connection $D_{h}$ (the Chern connection), compatible with the holomorphic structure of $E$ and the metric $h$. 
The curvature of this connection is given by $R_{h} =D_{h}\wedge D_{h}$, it is always a $(1,1)$ form with 
coefficients in ${\rm End}\,E$ and is called the Chern curvature. Using the Chern connection $D_{h}$ and the 
Higgs field $\phi$, one defines a new connection 
${\mathcal D}_{h} = D_{h} + \phi + \bar\phi_{h}$, where $\bar\phi_{h}$ is the adjoint of the Higgs field with 
respect to the metric $h$. The connection ${\mathcal D}_{h}$ and its curvature ${\mathcal R}_{h}={\mathcal D}_{h}\wedge{\mathcal D}_{h}$ 
are usually called the Hitchin-Simpson connection and curvature of $({\mathfrak E},h)$ and we say that the pair 
$({\mathfrak E},h)$ is Hermitian flat if ${\mathcal R}_{h}$ vanishes. The relation between the Hitchin-Simpson 
curvature and the Chern curvature is given by (see \cite{Bruzzo-Granha} or \cite{Cardona} 
for details)
\begin{equation}
 {\mathcal R}_{h} = R_{h} + D'_{h}(\phi) + D''(\bar\phi_{h}) + [\phi,\bar\phi_{h}]\,,  \label{Decomposition of R}
\end{equation}
where $D'_{h}$ and $D''$ are the holomorphic and antiholomorphic parts of the Chern connection and the 
commutator is an abbreviation for $\phi\wedge\bar\phi_{h} + \bar\phi_{h}\wedge\phi$. Consequently, 
the $(1,1)$ part of the Hitchin-Simpson curvature is 
\begin{equation}
 {\mathcal R}_{h}^{1,1} = R_{h} + [\phi,\bar\phi_{h}]\,, 
\end{equation}
and the mean curvature is defined, as usual, as the trace of ${\mathcal R}_{h}^{1,1}$ with respect to $g$. 
To be precise, if $\{z_{\alpha}\}_{\alpha=1}^{n}$ is a local coordinate system of the complex manifold $X$ and  
$\{e_{i}\}_{i=1}^{r}$ is a local frame field for ${\mathfrak E}$, with $\{e^{j}\}_{j=1}^{r}$ its dual frame, 
we have the following:  
\begin{equation}
  g=\sum g_{\alpha\bar\beta}\,dz^{\alpha}\otimes d{\bar z}^{\beta}\,, \quad\quad\quad  h=\sum h_{i\bar k}\,e^{i}\otimes {\bar e}^{k}\,,     \nonumber
\end{equation}
\begin{equation}
  {\mathcal R}_{h}^{1,1} = \sum\Omega^{i} _{\,j\alpha\bar\beta}\,e_{i}\otimes e^{j}\,dz^{\alpha}\wedge d{\bar z}^{\beta}\,.  \nonumber
\end{equation}
Let ${\mathcal K}^{i}_{\,j} = g^{\alpha\bar\beta}\Omega^{i} _{\,j\alpha\bar\beta}$, where $g^{\alpha\bar\beta}$ 
denotes the components of the inverse of the matrix associated to $g$, then the mean curvature of the Hitchin-Simpson 
connection is given by:
\begin{equation}
 {\mathcal K} =  \sum{\mathcal K}^{i}_{\,j}\,e_{i}\otimes e^{j}\,. \label{H-S mean End}
\end{equation}
It is important to note that the Hitchin-Simpson mean curvature is an endomorphism of $E$ which depends on 
$h$ and $g$ (we will not write this dependence explicitly in order to simplify the notation). Equivalently, 
by defining ${\mathcal K}_{j\bar k} = \sum h_{i\bar k}{\mathcal K}^{i}_{\,j}$ we can consider the mean curvature 
as the Hermitian form
\begin{equation}
 \hat{\mathcal K} = \sum{\mathcal K}_{j\bar k}\,e^{j}\otimes {\bar e}^{k}\,.  \label{H-S mean form}
\end{equation}
From the definitions (\ref{H-S mean End}) and (\ref{H-S mean form}) it is clear that the 
endomorphism ${\mathcal K}$  and the form $\hat{\mathcal K}$ are related by the metric $h$. To be precise, 
if $s$ and $t$ are sections of ${\mathfrak E}$, we have
\begin{equation}
 \hat{\mathcal K}(s,t) = h({\mathcal K}s,t)\,. \label{Relation HS mean curvatures}
\end{equation}

Now, associated to each Hermitian metric $g$ on $X$, there exists a fundamental $2$-form of type $(1,1)$,
also called the K\"ahler form of $X$ (see \cite{Siu} or \cite{Griffiths-Harris} for more details), which is defined by  
\begin{equation}
 {\omega} = i \sum g_{\alpha\bar\beta}\,dz^{\alpha}\wedge d{\bar z}^{\beta}\,. \nonumber
\end{equation}
If ${\omega}$ is closed, $g$ is called a K\"ahler metric of $X$ and the pair $(X,g)$ is called a K\"ahler 
manifold. Taking the wedge product of the fundamental form $n$-times we obtain the $(n,n)$ form
\begin{equation}
 {\omega}^{n} = i^{n}n!({\rm det\,}g)\,dz^{1}\wedge d{\bar z}^{1}\wedge\cdots\wedge dz^{n}\wedge d{\bar z}^{n}  \,. \nonumber
\end{equation}

We say that a Hermitian Higgs bundle $({\mathfrak E},h)$ satisfies the weak Hermitian-Yang-Mills condition or 
that it is weak Hermitian-Yang-Mills\footnote{It is 
also used the terminology weak Hermitian-Einstein or weak Hermitian-Yang-Mills-Higgs. We say also that 
$h$ is weak Hermitian-Yang-Mills.} if ${\mathcal K} = \gamma I$ 
for some function $\gamma$ (where $I$ is the identity endomorphism of $E$), or equivalently if 
$\hat{\mathcal K} = \gamma h$. Using standard identities for the curvature (see for instance \cite{Kobayashi} 
or \cite{Cardona}), we see that the property of being weak Hermitian-Yang-Mills is preserved under tensor 
products and dualization. To be more precise, $({\mathfrak E}^{*},h^{*})$ is weak Hermitian-Yang-Mills with 
factor $-\gamma$ if $({\mathfrak E},h)$ is weak Hermitian-Yang-Mills with factor $\gamma$, and 
$({\mathfrak E}_{1}\otimes{\mathfrak E}_{2},h_{1}\otimes h_{2})$ is weak Hermitian-Yang-Mills with factor 
$\gamma_{1} + \gamma_{2}$ if $({\mathfrak E}_{1},h_{1})$ and $({\mathfrak E}_{2},h_{2})$ are weak 
Hermitian-Yang-Mills with factors $\gamma_{1}$ and $\gamma_{2}$, respectively. Notice that in particular 
from this two results it follows that the Hermitian Higgs bundle 
$({\mathfrak E}^{\otimes p}\otimes{\mathfrak E}^{*\otimes q},h^{\otimes p}\otimes h^{*\otimes q})$ is weak 
Hermitian-Yang-Mills with factor $(p-q)\gamma$ if $({\mathfrak E},h)$ is Hermitian-Yang-Mills with factor $\gamma$. \\  

We say that a Hermitian Higgs bundle satisfies the Hermitian-Yang-Mills condition (or that it is 
Hermitian-Yang-Mills) if $\gamma$ is constant. Hermitian-Yang-Mills bundles have been studied in detail by Simpson \cite{Simpson}, \cite{Simpson 2} in the 
case of K\"ahler manifolds. Indeed, Simpson proved a Hitchin-Kobayashi correspondence for these bundles: a 
Hermitian Higgs bundle is Hermitian-Yang-Mills if and only if it is polystable. \\    

Let $({\mathfrak E},h)$ be a Hermitian Higgs bundle over $(X,g)$ and suppose $a=a(x)$ is a real 
positive function on $X$, then $h' = ah$ defines another Hermitian metric on ${\mathfrak E}$, i.e., 
$({\mathfrak E},h')$ is another Hermitian Higgs bundle. Since these two metrics are related by a conformal 
change, $\phi_{h'}=\phi_{h}$ (see \cite{Cardona} for details), and the corresponding Hitchin-Simpson mean 
curvatures are related by
\begin{equation}
 {\mathcal K}' = {\mathcal K} - \Box({\rm log}\,a)I\,, \label{Change HS mean c.}
\end{equation}
where $\Box = \sum g^{\alpha\bar\beta}\,\partial_{\alpha}\partial_{\bar\beta}$. This formula is indeed a 
generalization of a classical result for Chern mean curvatures, and it is important in the theory; for instance, from 
 (\ref{Change HS mean c.}) it follows that for every weak Hermitian-Yang-Mills metric $h$, there exists a conformal change $a$ such 
that $ah$ is Hermitian-Yang-Mills (see again \cite{Cardona} for details). Equivalently, we can rewrite the 
formula (\ref{Change HS mean c.}) in terms of Hermitian forms. Indeed, using (\ref{Relation HS mean curvatures}) we get
\begin{equation}
 \hat{\mathcal K}' = \hat{\mathcal K} - \Box({\rm log}\,a)h\,. \label{Change HS mean c. 2}
\end{equation}

On the other hand, for holomorphic sections of Hermitian vector bundles we have a formula involving the Chern 
curvature and the metric. If $d'$ and $d''$ denotes the $(1,0)$ and $(0,1)$ parts of $d$, and $s$ is a 
holomorphic section of the Hermitian vector bundle $(E,h)$, this formula is given by:
\begin{equation}
 d'd''h(s,s) = h(D's,D's) - h(Rs,s)\,, \label{Classical formula}
\end{equation}
where $D'$ denotes the $(1,0)$ part of the Chern connection $D$, and $R$ is the corresponding Chern 
curvature. In terms of local coordinates and local frame fields, $s=\sum s^{i}e_{i}$ and we can write (see \cite{Kobayashi} for more details)
\begin{equation}
 D's = \sum\nabla_{\alpha}s^{i}dz^{\alpha}e_{i}\,, \quad\quad\quad  d''s = \sum\nabla_{\bar\beta}s^{i}d{\bar z}^{\beta}e_{i} \,. \nonumber
\end{equation}
Therefore, the identity (\ref{Classical formula}) is equivalent to 
\begin{equation}
 \partial_{\alpha}\partial_{\bar\beta}\,h(s,s) = \sum h_{i\bar k}\nabla_{\alpha}s^{i}\,\nabla_{\bar\beta}{\bar s}^{k} 
                                               -  \sum h_{i\bar k}\,R^{i}_{\, j\alpha\bar\beta}\,s^{j}{\bar s}^{k} \,, \label{Classical formula loc.}
\end{equation}
where $R^{i}_{\, j\alpha\bar\beta}$ denotes here the components of the Chern curvature $R$. Now, by defining
\begin{equation}
  |D's|^{2} =  \sum h_{i\bar k}\,g^{\alpha\bar\beta}\,\nabla_{\alpha}s^{i}\,\nabla_{\bar\beta}{\bar s}^{k} \nonumber
\end{equation}
(the usual norm of $D'_{h}s$ with respect to $h$ and $g$) and taking the trace of the formula (\ref{Classical formula loc.}) 
with respect to the metric $g$, we obtain finally the Weitzenb\"ock formula:
\begin{equation}
 \Box h(s,s) = \lvert D's \lvert^{2} - \,{\hat K}(s,s)\,. \label{Weitzenbock formula}
\end{equation}
The Weitzenb\"ock formula plays an important role in the proof of the first vanishing theorem for holomorphic 
vector bundles, and as we will see, it is also important in the context of Higgs bundles.

\section{Vanishing theorems}

As we said before, in the study of holomorphic vector bundles one has some results on vanishing of holomorphic 
sections if some specific conditions on the curvature apply. These results, generically called Bochner's 
vanishing theorems, play an important role in complex geometry. Some of these vanishing results also holds 
in the context of Higgs bundles, in that case, we must replace the ordinary mean curvature by the 
Hitchin-Simpson curvature. \\

We establish here a first Bochner's vanishing theorem for Hermitian Higgs bundles over compact Hermitian 
manifolds. This theorem has been proved first by Bruzzo and Gra\~na-Otero \cite{Bruzzo-Granha} 
when the manifold was compact K\"ahler. However, it can be extended to non-K\"ahler manifolds. We write here 
the proof presented in \cite{Bruzzo-Granha} adapted to Higgs bundles over compact complex manifolds. \\

\begin{thm} \label{Main theorem} 
 Let $({\mathfrak E},h)$ be a Hermitian Higgs bundle over a compact Hermitian manifold $(X,g)$. Then\\ 
\noindent {\bf (i)} If the Hitchin-Simpson mean curvature $\hat{\cal K}$ is seminegative definite everywhere 
on $X$, then every $\phi$-invariant section $s$ of ${\mathfrak E}$ is parallel in the classical sense, 
i.e., $Ds=0$ with $D$ the Chern connection of $h$, and satisfies 
\begin{equation}
 \hat{\cal K}(s,s)=0\,.  \nonumber
\end{equation}
\noindent {\bf (ii)} If the Hitchin-Simpson mean curvature $\hat{\cal K}$ is seminegative definite 
everywhere on $X$ and negative definite at some point of $X$, then ${\mathfrak E}$ has no nonzero 
$\phi$-invariant sections. 
\end{thm}

\noindent {\it Proof:} Let $s$ be a $\phi$-invariant section of ${\mathfrak E}$ and assume 
$\hat{\cal K}$ is seminegative definite everywhere. From the decomposition (\ref{Decomposition of R}) of 
the Hitchin-Simpson curvature we have
\begin{equation}
 {\mathcal R}s  =  Rs  +  D'(\phi)s  +  D''(\bar\phi)s  + [\phi,\bar\phi]s\,. \nonumber
\end{equation}
Since $s$ is $\phi$-invariant $[\phi,\bar\phi]s=0$ and hence, taking the trace with respect to $g$ 
in the above expression we get ${\cal K}s = Ks$, or equivalently $\hat{\cal K}(s,s)=\hat{K}(s,s)$. 
Then, using the classical Weitzenb\"ock formula (\ref{Weitzenbock formula}) we obtain
\begin{equation}
 \Box h(s,s) = \lvert D's\rvert^{2} - \hat{\cal K}(s,s)\,.  \label{Weitzenbock formula Higgs}
\end{equation}
Now, since $\hat{\cal K}$ is seminegative definite, the right hand side of 
(\ref{Weitzenbock formula Higgs}) is non-negative and by Hopf's maximum principle 
(see e.g. \cite{Kobayashi} or \cite{Lubke}) this implies that $h(s,s)$ is constant and consequently  
$\Box h(s,s) = 0$. Therefore, necessarily $\hat{\cal K}(s,s)=0$ and $D's=0$, but since $s$ is holomorphic 
this last condition is equivalent to $Ds=0$ and (i) follows. \\

On the other hand, suppose now $s$ is a $\phi$-invariant section of $({\mathfrak E},h)$ and assume this time 
that $\hat{\cal K}$ is seminegative definite everywhere and negative at some point. Then, from (i) we know
that $s$ is parallel with respect to the Chern connection and hence it never vanishes. Using again (i) we have
$\hat{\cal K}(s,s)=0$ and we have a contradiction, because $\hat{\cal K}$ must be negative at some point of 
$X$. \; Q.E.D. \\

As in the classical case, from this first vanishing theorem we have other results involving tensor 
products. In particular we have the following
\begin{thm}
 Let $({\mathfrak E}_{1},h_{1})$ and $({\mathfrak E}_{2},h_{2})$ be two Hermitian Higgs bundles over a 
 compact Hermitian manifold $(X,g)$ and let $\hat{\cal K}_{1}$ and $\hat{\cal K}_{2}$ be the corresponding 
 Hitchin-Simpson mean curvatures. Let $\phi$ be the Higgs field of $({\mathfrak E}_{1}\otimes{\mathfrak E}_{2},h_{1}\otimes h_{2})$ 
 and let $\hat{\cal K}_{1\otimes2}$ be the Hitchin-Simpson mean curvature of this tensor product. Then \\
 \noindent {\bf (i)} If both $\hat{\cal K}_{1}$ and $\hat{\cal K}_{2}$ are seminegative definite everywhere 
 on $X$, then every $\phi$-invariant section $\xi$ of 
 ${\mathfrak E}_{1}\otimes{\mathfrak E}_{2}$ is parallel with respect to the 
 Chern connection $D_{1\otimes2}$ (induced from $D_{1}$ and $D_{2}$), i.e, $D_{1\otimes2}\xi =0$, and 
 satisfies
 \begin{equation}
 \hat{\cal K}_{1\otimes2}(\xi,\xi) = 0\,.  \nonumber
 \end{equation}
 \noindent {\bf (ii)} If both $\hat{\cal K}_{1}$ and $\hat{\cal K}_{2}$ are seminegative definite everywhere 
 on $X$ and either one is negative definite at some point of $X$, then 
 ${\mathfrak E}_{1}\otimes{\mathfrak E}_{2}$ admits no nonzero $\phi$-invariant sections.
\end{thm}

\noindent {\it Proof:} From \cite{Cardona} we know the Hitchin-Simpson mean curvature of the Hermitian Higgs 
bundle $({\mathfrak E}_{1}\otimes{\mathfrak E}_{2},h_{1}\otimes h_{2})$ satisfies
\begin{equation}
 {\cal K}_{1\otimes2} =  {\cal K}_{1}\otimes I_{2} + I_{1}\otimes{\cal K}_{2}\,. \nonumber
\end{equation}
Now, similarly to classical case, by choosing orthonormal local frame fields we can represent locally  
${\cal K}_{1}$ and ${\cal K}_{2}$ by diagonal matrices, and hence ${\cal K}_{1\otimes2}$ becomes also a 
diagonal matrix whose nonzero elements are sums of the diagonal elements of ${\cal K}_{1}$ and 
${\cal K}_{2}$. That is, if $a_{i}$ and $b_{j}$ are the diagonal elements of ${\cal K}_{1}$ and 
${\cal K}_{2}$, the diagonal elements of ${\cal K}_{1\otimes2}$ are $a_{i} + b_{j}$. At this point, since 
the local frame field is orthonormal, ${\mathcal K}_{j\bar k} = \sum \delta_{i\bar k}{\mathcal K}^{i}_{\,j} = {\mathcal K}^{k}_{\,j}$ and (i) and (ii) follow from 
Theorem \ref{Main theorem}.  \; Q.E.D. \\
\begin{cor}
 Let $({\mathfrak E},h)$ be a Hermitian Higgs bundle over a compact Hermitian manifold $(X,g)$. Let 
 $({\mathfrak E}^{\otimes p},h^{\otimes p})$ be the tensor product of $({\mathfrak E},h)$ $p$-times and let 
 $\psi$ be its Higgs field (constructed from the Higgs field $\phi$ of ${\mathfrak E}$) and $\hat{\cal K}_{\otimes p}$ its 
 Hitchin-Simpson mean curvature. Then \\
 \noindent {\bf (i)} If $\hat{\cal K}$ is seminegative definite everywhere on $X$, then every $\psi$-invariant 
 section $\xi$ of ${\mathfrak E}^{\otimes p}$ is parallel in the classical sense, i.e.,  
 $D_{\otimes p}\xi = 0$, and satisfies
 \begin{equation}
 \hat{\cal K}_{\otimes p}(\xi,\xi) = 0\,.  \nonumber
 \end{equation}
 \noindent {\bf (ii)} If $\hat{\cal K}$ is seminegative definite everywhere on $X$ and negative definite 
 at some point of $X$, then ${\mathfrak E}^{\otimes p}$ admits no nonzero $\psi$-invariant 
 sections. 
\end{cor}

Let $TX$ and $T^{*}X$ be the tangent and cotangent bundles to $X$, then we can consider the pairs $(TX,g)$ 
and $(T^{*}X,g^{*})$ as Hermitian Higgs bundles with zero Higgs fields. Since the Hermitian metrics $h$ 
and $g^{*}$ on $E$ and $T^{*}X$ induce a Hermitian metric, say $k$, on the Higgs bundle
$\Omega^{p}({\mathfrak E}) = {\mathfrak E}\otimes\bigwedge^{p} T^{*}X$, the pair 
$(\Omega^{p}({\mathfrak E}),k)$ can be considered as a Hermitian Higgs bundle over $(X,g)$. We denote by 
${\hat K}_{TX}$ the mean curvature\footnote{Notice that the Higgs field of $(TX,g)$ is zero, hence the
Hitchin-Simpson mean curvature and the Chern mean curvature coincide for this Hermitian Higgs bundle.} 
of $(TX,g)$ and by $\hat{\cal K}_{\Omega^{p}}$ the Hitchin-Simpson curvature of 
$(\Omega^{p}({\mathfrak E}),k)$. From this we have the following
\begin{cor}
 Let $({\mathfrak E},h)$ be a Hermitian Higgs bundle over a compact Hermitian manifold $(X,g)$ with Higgs field $\phi$, 
 and denote by $\hat{\cal K}_{E}$ its Hitchin-Simpson mean curvature. Then \\
  \noindent {\bf (i)} If $\hat{\cal K}_{E}$ is seminegative definite and $\hat{K}_{TX}$ is semipositive 
  definite everywhere on $X$, then every $\phi$-invariant section $\xi$ of $\Omega^{p}({\mathfrak E})$ 
  is parallel in the classical sense and satisfies
  \begin{equation}
 \hat{\cal K}_{\Omega^{p}}(\xi,\xi) = 0\,.  \nonumber
 \end{equation}
 \noindent {\bf (ii)} If $\hat{\cal K}_{E}$ is seminegative definite and $\hat{K}_{TX}$ is semipositive 
  definite everywhere on $X$, and either $\hat{\cal K}_{E}$ is negative definite or $\hat{K}_{TX}$ is 
  positive definite at some point of $X$, then $\Omega^{p}({\mathfrak E})$ admits no nonzero $\phi$-invariant 
  sections.  
\end{cor}

On the other hand, as it is well known (see \cite{Kobayashi} for details), there exists vanishing 
theorems for Hermitian vector bundles if certain condition for the greatest eigenvalue of the Chern mean 
curvature holds, and for Hermitian vector bundles satisfying the Hermitian-Einstein condition. These results can 
be extended to Hermitian Higgs bundles if we use the Hitchin-Simpson mean curvature instead of the Chern mean 
curvature. To be precise we have the following results
\begin{thm}
 Let $({\mathfrak E},h)$ be a Hermitian Higgs bundle over a compact Hermitian manifold $(X,g)$ with fundamental 
 $2$-form $\omega$. Let $\lambda_{1}\leq\cdots\leq\lambda_{r}$ be the eigenvalues of the Hitchin-Simpson mean 
 curvature ${\cal K}$ of $({\mathfrak E},h)$. If 
 \begin{equation}
 \int_{X}\lambda_{r}\omega^{n} < 0\,,  \nonumber
 \end{equation}
 then ${\mathfrak E}$ admits no nonzero $\phi$-invariant sections.
\end{thm}

\noindent {\it Proof:} In a similar way to the proof of the classical case, we consider a $C^{\infty}$ function 
 $f$ on $X$ such that $\lambda_{r} < f$ and  
\begin{equation}
 \int_{X} f \omega^{n} = 0\,.  \nonumber
 \end{equation}
Clearly $f$ is orthogonal to all constant functions, but since $X$ is compact, every $\Box$-harmonic 
function is constant and hence $f$ is orthogonal to the $\Box$-harmonic functions. Therefore, we can consider 
a $C^{\infty}$ solution $u$ of the equation $\Box u = f$. Now, if we define the positive function $a = e^{u}$ 
and consider the Hermitian Higgs bundle $({\mathfrak E},h')$ where $h' = ah$, then 
$\Box({\rm log}\,a) = f$ and using (\ref{Change HS mean c.}) we have that ${\cal K}'$ is diagonal and 
negative definite. Consequently, also $\hat{\cal K}'$ is negative definite and the result follows from 
Theorem \ref{Main theorem}. \; Q.E.D. 

\begin{thm}
 Let $({\mathfrak E},h)$ be a Hermitian Higgs bundle over a compact Hermitian manifold $(X,g)$ and let 
 $\psi$ be the Higgs field of the Hermitian Higgs bundle 
 $({\mathfrak E}^{\otimes p}\otimes{\mathfrak E}^{*\otimes p},h^{\otimes p}\otimes h^{*\otimes p})$. If 
 $({\mathfrak E},h)$ satisfies the weak Hermitian-Yang-Mills condition, then every $\psi$-invariant section 
 of ${\mathfrak E}^{\otimes p}\otimes{\mathfrak E}^{*\otimes p}$ is parallel in the classical sense.
\end{thm}

\noindent {\it Proof:} Let $({\mathfrak E},h)$ be weak Hermitian-Yang-Mills with factor $\gamma$, then from 
results of the second section we know that 
$({\mathfrak E}^{\otimes p}\otimes{\mathfrak E}^{*\otimes p},h^{\otimes p}\otimes h^{*\otimes p})$ is 
Hermitian-Yang-Mills with factor $(p-p)\gamma = 0$ and hence the mean curvature vanishes identically. 
At this point the result follows from Theorem \ref{Main theorem}. \; Q.E.D. \\

\noindent{\bf Acknowledgements} \\

\noindent The author would like to thank Conacyt and SNI for partial support and in particular to 
U. Bruzzo and R. Herrera for some comments and suggestions. The author would like to thank also 
G. Dossena, O. Mata and C. Mar\'in.

\end{document}